\documentclass[prd,aps,floats,preprintnumbers,preprint,
12pt, showpacs]{revtex4}
\usepackage{epsfig}

\begin{document}
\preprint{MCTP-04-32}


\title{Momentum Representation of Coulomb Wave Functions and Level 
Shifts in Bottomonium due to Charm Effects}

\author{Haibin Wang}
\email[]{haibinw@physics.purdue.edu}
\affiliation{Department of Physics, Purdue University, West Lafayette,
IN 47907, USA}

\author{York-Peng Yao}
\email[]{yyao@umich.edu}
\affiliation{Michigan Center for Theoretical Physics, University of Michigan, 
Ann Arbor, MI 48109, USA}

\begin{abstract}
Since effective potentials derived from Feynman diagrams are naturally 
given in momentum space, we formulate the non-relativistic Coulomb problem 
entirely in momentum representation.  We give momentum wave functions 
for all quantum numbers in one-dimensional integrals, even though they 
can be evaluated.  Angular momentum decomposed Green's functions are
then compactly represented.  We apply this formalism to investigate
the next to next leading order charm effects on 1S bottomonium level shift.  
Our one insertion results are given completely in analytic form and 
numerically agree with previous results.
Our two insertion results are also in agreement.
The net effect of finite charm mass
is to decrease the bottom mass by 33 MeV, as determined through the measured
1S energy.
\end{abstract}

\maketitle

\newpage
\section{Introduction}

It is a common belief that new physics most likely will be due to new dynamics at a higher scale.  
Upon accepting this premise, we should expect that bigger effects will appear in processes
in which known heavy particles, such as the bottom quark, the top quark and the
Higgs boson 
actively participate.  Such interplay between scales can be quite intricate, as exemplified
in the $\rho$-parameter and in quite a few rare B decays.  We would like on the one hand
to use some of these processes to determine the masses of the known heavy particles as best as 
we can because the rates of many interesting processes depend on them strongly and on the 
other to lay the ground work to look for discrepancies as a byway to new physics.  It is incumbent 
on us that for such processes the known dynamics must be well understood and that the 
so far 
uncalculable effects be well accounted for or at least be under control.

\bigskip
One active area is the first few of bottomonium energy levels $[1]$ and the production threshold of 
$t\bar t$ $[2]$.  Here as the starting approximation, the heavy quarks move under a color Coulomb
potential non-relativistically.  Because retardation is small, one can add other effective 
interaction pieces as perturbations by integrating out the fast degrees of freedom.  
One can then extract out the bottom mass from the first couple of measured energy levels.
For the top quark, it decays too fast to have real bound states, but these would-be bound
states have strong effects on the shape of the production cross section near the 
thresholds.

\bigskip
In the zeroth order approximation, such systems are just like a hydrogen atom.  In most
of the treatments of quarkonium systems, Coulomb-like wave functions
in the coordinate space are used to evaluate matrix elements. 
We would like to ask whether an alternative
may be also viable, for the reason that when one treats such systems in conjunction with
quantum field theory, most if not all radiative correction calculations are done in 
momentum space via Feynman diagrams at some stage.  In an effective Lagrangian approach,
the Wilson coefficients are calculated in momentum space, but the operators are customarily
converted back into coordinate space and their matrix elements are then obtained by 
averaging with spatial wave functions.  It is therefore of some interest to see how to bypass the 
last step by following the momentum approach throughout.

\bigskip
Needless to say, hydrogen wave functions in momentum representation have been repeatedly used $[3]$
for a long time.  One difference we want to make here is that we have a convenient and compact
representation for them in one-dimensional integrals, which we shall keep in such a form
even though they can be evaluated into Gegenbauer polynomials.  Matrix elements can then
be evaluated mostly by the use of the method of residues and the result will collapse 
into only a small number of terms.  We shall also give a very simple representation for 
the propagators.

\bigskip
As we indicated earlier, we find it intriguing to explore the interplay of masses.
Therefore, we shall use as an example to apply the momentum technique to investigate
 the charm mass effects $[4-8]$ on the bottomonium level shifts to the next-next leading order.
One notices that the binding momentum $\approx {4\over 3}\alpha_s m_b \approx 2 \ GeV $
is comparable to $m_c \approx 1.25 \ GeV$.  One must take the charm loop as a whole as 
a term in the potential, which becomes highly non-trivial in the coordinate space.

\bigskip
There are two sets of diagrams.  The first set is due to
the insertions of the lowest order charm loop once or twice, or the limiting zero mass particle
effects and a charm loop each once on the same gluon line ( Fig.(1)).  
\begin{figure}
\center
\includegraphics[angle=-90,scale=0.85]{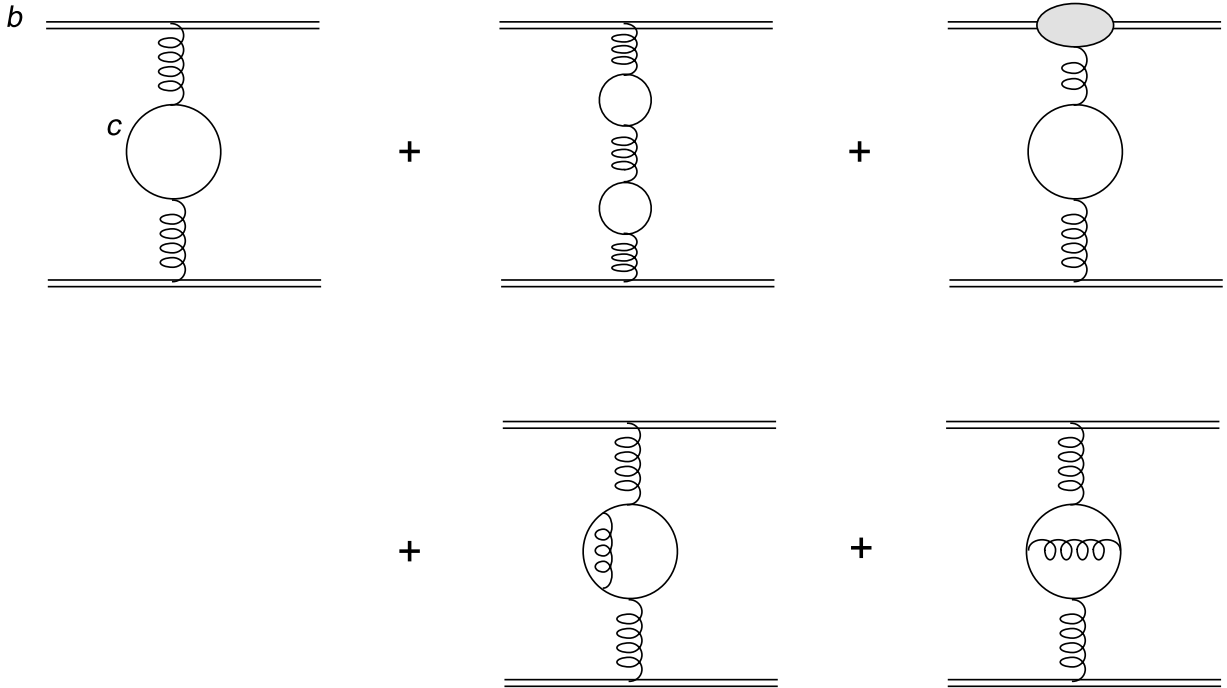}
\caption{We show schematically the next leading order and next-to-next leading order
charm mass effects in single insertions which alter the bottomonium energy levels.
The loops are composed of a charm and anti-charm pair, the shaded ellipses represent
cumulative second order zero mass quark and gluon effects, and the double lines
depict either a $b$ or $\bar{b}$ quark.}
\end{figure}
Also, there are diagrams
due to the fourth order charm loop.  (We neglect the charm 
loop effects on the vertex.)  We have been able to evaluate all of them analytically 
into simple functions.  They account for $98\%$ of the charm effects for the 1S level 
shift.  
The second set is due to double insertions of the above potential on two separate gluon 
lines ( Fig.(2) ) 
\begin{figure}
\center
\epsfig{file=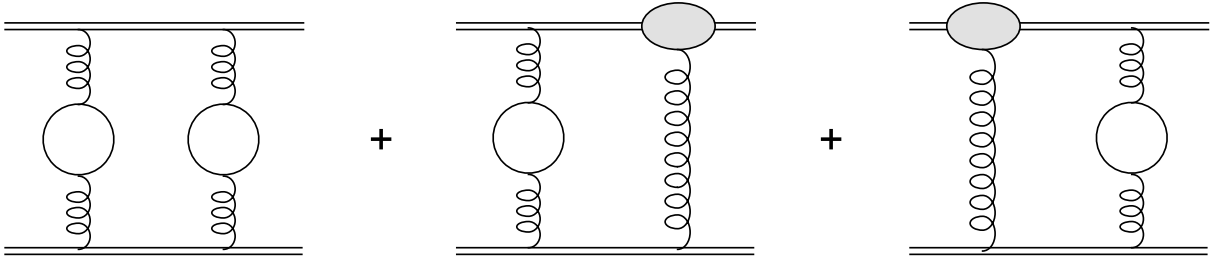,angle=-90,scale=0.85}
\caption{We show schematically the next-to-next leading order charm effects in
double insertions which alter the bottomonium energy levels. 
The loops are composed of a charm and anti-charm pair, the shaded ellipses represent
cumulative second order zero mass quark and gluon effects, and the double lines
depict either a $b$ or $\bar{b}$ quark.}
\end{figure}
and 
so far can be given only in terms of  some simple integrals.  
The results are in agreement with those
given in \cite{Hoang3}. 
Furthermore, because we have the explicit 
functions for the first set at hand, we can follow the relevant 
branch of the functions and continue them to cover the toponium would-be bound state
energy level shifts, here due to the bottom quark mass effects.  We hope that this example
is convincing enough to illustrate that the same technique can be used to cover exotic 
particle effects, when called upon. 

\bigskip
In an article \cite{Akhoury} co-authored by one of us, 
we showed how the technique presented here could be extended to yield results at least for the 
spherically symmetrical states of the hydrogen atom and gave a bound on the non-commutative 
scale in a certain extension of the non-relativistic kinematics 
\cite{Maggiore}, by using the highly accurate 1S-2S energy difference.

\bigskip
The plan of this paper is as follows: in the next section, we shall solve for the momentum wave
functions.  An operator algebra \cite{Ivash} will be used to normalize them.  These wave functions in one
dimensional integral representation will be the basis for our calculations later.  In section 3, 
we shall briefly display the Green's functions for arbitrary angular momentum, but the 
complete treatment is devoted to the S state, as we shall use it in a subsequent section.  
Our momentum wave functions will be used in section 4A to obtain the ground state energy 
level shifts for 1S bottomonium due to next leading and next to next leading order 
charm mass effects in single insertions.  We 
are able to give our results completely analytically in simple functions.  They agree with 
what were obtained partially analytically by others.  In section 4B, we give our formulation
for double insertions and carefully state our subtraction procedure.  A simple test is to 
apply it to obtain the ground state energy level of the modified Coulomb potential
$-{4\over 3}\alpha_s(1+\delta){1\over r}$, the exact result of which is known a priori.
Numerical analysis is presented in section 5 and some concluding remarks are made in 
section 6.  In an appendix, we further illustrate our formalism by evaluating some 
well known matrix elements of $({1\over r})^{0,1,2}$.


\section{Momentum Wave Functions}
  
In this section, we shall derive a one-dimensional representation for 
the color Coulomb wave functions and show that, when converted 
back into coordinate space, they are the same  as in textbooks. 
In fact, they will be slightly more general,
because in some cases where the perturbing potential has power 
law dependence on the radial 
coordinate, we can take care of it easily.  We shall 
use an algebra to determine the normalization factors 

From the Schrodinger equation, with the reduced mass $m=m_b/2$ for bottomonium,

\begin{equation}[E-{1\over 2m}({1\over i}{\vec \partial \over \partial \vec \xi})^2
+{4\over 3}\alpha_s{1\over \sqrt {\vec \xi^2}} ]\psi(\vec \xi)=0,
\label{eq2.1}
\end{equation}
we take away the centrifugal barrier and  perform  an angular momentum decomposition
\begin{equation}\psi(\vec \xi)=Y_{l,m }(\theta. \phi)\tilde\gamma_l(\xi)\xi^ {-(l+1)},
\end{equation}
to obtain 

\begin{equation}
[(E-{1\over 2m}({1\over i}{d\over d \xi})^2)\xi-{i \over m}(l+1)
({1 \over i }{d \over d \xi})+{4\over 3} \alpha_s]\tilde \gamma _l(\xi)=0.
\nonumber
\end{equation}
We have used $\vec \xi $ to denote coordinates, with $\xi$ as the radial distance and 
$\theta , \phi$ as the angles.  Now we perform a Fourier transform with respect to 
the {\it radial distance $\xi$}

\begin{equation}
\tilde\gamma_l(\xi)={1\over \sqrt{2\pi }}\int _{-\infty} ^\infty dpe^{ip\xi} 
\gamma_l (p)
\end{equation}
to arrive at \cite{Schwinger} 
\begin{equation}
[(E-{p^2\over 2m})i {d\over dp} -{i \over m}(l+1)p+{4\over 3} \alpha_s]
\gamma_l(p)=0.
\end{equation}

Let us denote the energy of the state as 
\begin{equation}
E={p_0^2\over 2m},
\end{equation}
and rearrange the equation into

\begin{equation}
({d\over dp}+{l+1+i\eta \over p-p_0}+{l+1-i\eta \over p+p_0})\gamma_l=0,
\end{equation}
where we have written

\begin{equation}
\eta ={4\over 3}\alpha_s{m \over p_0} .
\end{equation} 

The solution of this equation is 

\begin{equation}
\gamma_l(p)=A_l{1\over (p-p_0)^ {l+1+i\eta}}{1\over (p+p_0)^{l+1-i\eta}},
\end{equation}
where $A_l$ is an integration constant to be determined shortly by normalization.
The bound states are determined by the poles in the upper p-plane.  Thus 
for the negative energy solutions we set 
\begin{equation}
p_0=i\kappa, \  \ \eta =-in,
\end{equation}
in which we require

\begin{equation}
n-l-1= r=0,1,2...
\end{equation}
so that there is no pole in the lower p-plane.  We do not want to burden our expressions
in the remainder of this section with indices and therefore it is understood that we shall
deal with states with the same principal quantum number n one at a time.

\bigskip
To determine $A_l$, we go back to the coordinate representation, where we notice
that  

\begin{equation}
u_l(\xi)\equiv\xi^{-l} \tilde \gamma_l(\xi)
\end{equation}
satisfies the equation
\begin{equation}
H_lu_l=Eu_l,
\end{equation}
where 
\begin{equation}
H_l\equiv {p^2\over 2m}+{l(l+1)\over 2m\xi^2}-{4\over 3}\alpha_s{1\over \xi}, \ \ p\equiv {1\over i }
{d\over d\xi}.
\end{equation}
 
Let us consider the operator \cite{Ivash}
\begin{equation}
L_l\equiv p-i({ l\over \xi}-{4\over 3}\alpha_s{m\over  l}).
\end{equation}
Some straightforward algebra shows that we have

\begin{equation}
L_lH_l-H_{l-1}L_l=0.
\end{equation}
It follows that if $u_l$ is an eigenvector of $H_l$, so is $L_lu_l$ of $H_{l-1}$, 
i.e. if the eigenenergy is $E_0$, 
\begin{equation}  
H_lu_l= E_0 u_l, \nonumber
\end{equation}
then
\begin{equation}
L_lH_lu_l=H_{l-1}L_lu_l ,  \rightarrow E_0(L_lu_l)=H_{l-1}(L_lu_l). \nonumber
\end{equation}
$E_0$ as well known depends only on n but not l for bound states.  
One can easily arrive at
 
\begin{equation}
{1\over 2m}L_l^\dagger L_l=H_l+ ({4\over 3}\alpha_s)^2{m\over 2l^2}.
\end{equation}
Therefore, for a given n, $L_l$ acts as a lowering operator

\begin{equation}
L_lu_l=B_lu_{l-1},
\end{equation}
which gives a norm
 
\begin{equation}
{1\over 2m}|B_l|^2=({4\over 3}\alpha_s)^2{m\over 2}({1\over l^2}-{1\over n^2}),
\end{equation}
upon using the bound state energies

\begin{equation}
E_0=-({4\over 3}\alpha_s)^2{m\over 2n^2}.
\end{equation}

We shall follow the conventional choice to make the wave functions $u_l$ real, which    
dictates the choice of the phase for $B_l$ so that

\begin{equation}
L_lu_l=-i{1 \over a}\sqrt {{1\over l^2}-{1\over n^2}} u_{l-1}, 
\ a^{-1}\equiv {4\over 3}\alpha_s  m.
\end{equation}
When we transcribe this last equation into the momentum representation, we have

\begin{equation}
{1 \over a}\sqrt {{1\over l^2}-{1\over n^2}}i {d\over dp} \gamma_{l-1}=(ip -
{1\over la})\gamma_l.
\end{equation}
Upon using the explicit solutions for $\gamma(p)$'s, we arrive at 

\begin{equation}
A_l=-{2\over na}\sqrt{n^2-l^2}A_{l-1}, \nonumber
\end{equation}
and upon iteration

\begin{equation}
A_l=(-2 \kappa)^l\sqrt {{(n+l)!\over n(n-l-1)!}}A_0.
\end{equation}
We have used the relation
\begin{equation}
\kappa={1 \over na},
\end{equation}
and in conformity with the standard choice of phase, we  fix
\begin{equation}
A_0=-\sqrt{{2\kappa^3\over \pi}}.
\end{equation}

\bigskip
At this point we want to be reminded that the radial wave functions are 
$\tilde \gamma_l(\xi)\xi^{-(l+1)}$.  Because we shall use them to evaluate matrix 
elements of operators which may have $\xi$ dependence, let us define more 
generally

\begin{equation}
{\tilde \gamma _l(\xi)\over \xi^t}\equiv \tilde \gamma ^t_l(\xi),
\end{equation}
the Fourier transform of which 
\begin{equation}
\tilde\gamma_l^t(\xi)={1\over  \sqrt{2\pi }}\int_{-\infty}^\infty
dp \ e^{ip\xi} \ \gamma_l^t(p)
\label{eq2.26}
\end{equation}
satisfies
\begin{equation}
i {d\over dp}\gamma_l^t(p)=\gamma_l^{t-1}(p),
\end{equation}
and therefore
\begin{equation}
(i {d\over dp})^t\gamma_l^t(p)=\gamma_l^0(p)\equiv \gamma_l(p).
\end{equation}
It can be inverted to yield

\begin{equation}
(i)^t\gamma_l^t(p)=(-1)^{t+1}{1\over (t-1)!}\int_{-\infty}^0
dp'\ p'^{t-1}A_l{(p'+p+p_0)^{n-l-1}\over (p'+p-p_0)^{n+l+1}} \ ,
\end{equation}
for a given n and $t\ge1$.  This is the basic integral representation we shall use, 
although it can be evaluated into Gegenbauer polynomials.  

\bigskip
To proceed further, we want to write Laguerre polynomials $L_r^k$ in an 
integral representation over momentum.  From the definition of 
$\tilde \gamma (\xi)$ and how $L_r^k$ appears in the wave function, i.e.

\begin{eqnarray}
\psi _{nlm} &=& Y_{lm}(\theta, \phi){\tilde \gamma _l \over \xi^{l+1}}
\nonumber \\
 &=& Y_{lm}(\theta, \phi)\sqrt{(2\kappa )^3
{(n-l-1)!\over 2n(n+l)!}}e^{-\kappa \xi}
(2\kappa \xi )^lL_r^k(2\kappa \xi ),
\end{eqnarray}
where $r=n-l-1$ and $k=2l+1$, we have
\begin{eqnarray}
L_r^k(\rho\equiv 2\kappa \xi ) &=& -{(r+k)!\over r!}
{1\over (k-1)!}e^{\kappa \xi}\int _{-\infty}^0dp'p'^{k-1} \nonumber \\
& &
\times {1\over 2\pi i}\int_{-\infty}^\infty dp e^{ip\xi}{(p+p'+p_0)^r\over 
(p+p'-p_0)^{r+k+1}}.
\label{eq2.32}
\end{eqnarray}

One can of course carry out the p and p' integrations to obtain

\begin{equation}
L_r^k(\rho)={e^\rho\over r!}\rho^{-k} ({d\over d\rho})^r \rho^{r+k} e^{-\rho}.
\end{equation}
However, for some problem where a sum over r has to be performed,
it is more useful to have $L_r^k$ presented as in Eq. (\ref{eq2.32}).  
\bigskip

Before we leave this section, we give a formula which is useful in the calculation of 
matrix elements,

\begin{equation}
\int_0^\infty dp_1 \ dp_2 \ {p_1^i p_2^j\over (p_1+p_2+b)^k}
={\Gamma(i+1)\Gamma(j+1)\over \Gamma(i+j+2)}\int_0^\infty dp'{p'^{i+j+1}
\over (p'+b)^k} \ ,
\label{eq2.34}
\end{equation}
when $k>i+j+2.$  In an appendix, we shall use it to reproduce some of the 
classic matrix elements to illustrate further our formalism.

\section{Propagators}
\bigskip
In order to carry out double insertions for the next-next leading order
energy shifts, we need the propagators which 
are defined in

\begin{equation}
G(\vec x,\vec y;E)\equiv -\sum_i{|i><i| \over E-E_i}
=\sum_l(2l+1)G_l(x,y;E)P_l({\vec x \cdot \vec y\over xy}),
\end{equation}
where $P_l$ are the Legendre polynomials and please note the 
minus sign in the sum over the eigen-energies $E_i$.  We have assumed a 
central 
(Coulomb) potential in writing down the above expression.  Because of an 
interesting result in angular momentum decomposed Green's functions
\cite{Hostler}, 
we make a slight change of notation and write the energy as 
\begin{equation}
E=-\kappa_0^2/m_b=p_0^2/m_b,
\end{equation}
then 

\begin{eqnarray}
G_l(x,y; E)&=&{m_b\kappa_0\over 2\pi}(2\kappa_ox)^l(2\kappa_0 y)^l
e^{-\kappa_0(x+y)}\sum_{r=0}^\infty {L_r^{2l+1}(2\kappa_0x)
L_r^{2l+1}(2\kappa_0y)r!\over (r+l+1-\nu)(r+2l+1)!} \nonumber \\
&=&{m_b\kappa_0\over 2\pi}\sum _{n=1}^\infty {2n\over (2\kappa_0)^3}
{1\over n-\nu}{u_n^l(2\kappa_0x)\over x}{u_n^l(2\kappa_0y)\over y},
\label{eq3.3}
\end{eqnarray}
where we have restored the principal quantum number n to label the radial
wave functions (i.e. $u_l\rightarrow u_n^l) $ and defined
\begin{equation}
\nu\equiv {m_b\over 2\kappa_o}({4\over 3}\alpha_s).
\end{equation}
We shall later on be interested in the behavior $\nu \rightarrow 1$ when we look 
into the ground state.  
Now because of Eq.(\ref{eq2.32}), which gives the reduced radial functions

\begin{eqnarray}
{u_n^l(2\kappa_0x)\over x}&=&(i)^{-(l+1)}(-1)^l{1\over l!}A_l 
({1\over 2\pi})^{1/2}
\nonumber \\
& & \times \int_{-\infty}^\infty dp_1e^{ip_1x}\int_{-\infty}^0dp_1' \ p_1'^l \ 
{(p_1'+p_1+p_0)^{n-l-1}\over (p_1'+p_1-p_0)^{n+l+1}},
\end{eqnarray}
and a similar expression for ${u_n^l(2\kappa_0y)\over y}$, with $p_1\to p_2$
and $p_1'\to p_2'$.  Also, we write
\begin{equation}
{1\over n-\nu}=\int _0^1d\rho \rho^{n-\nu-1},
\end{equation}
assuming $1-\nu  >0.$ 
Performing the sum over n in Eq.(\ref{eq3.3}), we find 
\begin{eqnarray}
G_l(x,y;E) &=& (-1)^l {m_b(2\kappa_0)^k\over 4\pi}{k!\over (l!)^2}({1\over 2\pi i})^2
\int_{-\infty}^\infty dp_1 e^{ip_1x}\int_{-\infty}^\infty dp_2 e^{ip_2y}
\nonumber \\ 
 & &
\times  \int_{-\infty}^0dp_1'p_1'^l\int_{-\infty}^0dp_2'p_2'^l\int _0^1 d\rho \rho^{l-\nu} 
\nonumber \\ 
& & \times{1\over \big( (p_1+p_1'-p_0)
(p_2+p_2'-p_0)-\rho (p_1+p_1'+p_0)(p_2+p_2'+p_0)\big)^{k+1}},
\nonumber \\
 & & k\equiv 2l+1.
\end{eqnarray}
For the case $l=0$, it is 
easy to perform the sum and integrate over $p_1', p_2' $ to obtain

\begin{eqnarray}
G_{l=0}(x,y;E)&=&{-m_b\over 8\pi \kappa_0}({1\over 2\pi i})^2 
\int_{-\infty}^\infty dp_1 e^{ip_1x}\int_{-\infty}^\infty dp_2 e^{ip_2y}
\nonumber \\
& & \times \int_0^1 d\rho \rho^{-1-\nu}ln({bc \over ad}),
\label{eq3.8}
\end{eqnarray}
where 
\begin{equation}
bc= (p_1-p_0)(p_2-p_0)-\rho (2p_1p_2-2p_0^2)+\rho^2( p_1+p_0)(p_2+p_0),
\end{equation}
and 
\begin{equation}
ad=bc-4\rho p_0^2.
\end{equation} 

With the wave functions and the propagators, we can proceed  to calculate energy shifts in the 
next sections.

\section{Energy Shifts due to Charm Effects}

One objective in high precision tests of fundamental physics 
is to give operational meaning to the parameters in a theory and to 
measure them.   Among the many parameters in the Standard Model,
great improvements are being made in extracting the values of
heavy particle masses, both experimentally and theoretically.  For example, 
for the b-quark mass one way to determine it is by using the ground state 
of the bottomonium.  As for the top quark, although its life time is too short
for forming toponium, the location and the shape of the $ t \bar t$ 
threshold will yield crucial information.

Such considerations have been augmented to a very high and 
sophisticated degree by many groups.  Thus, the ground state 
energy level of the bottomonium has been calculated to an accuracy of 
$\alpha_s^5(ln(\alpha_s))$ [1].  At this order, among other contributions, 
there are the shifts due to the charm quark vacuum polarizations in the 
potential function.  The important ratio here for the energy shifts 
is $k/m_c$, where k is 
the momentum transfer, which is $\sim {2\over 3}\alpha_s m_b$.
Thus $k/m_c \approx 1$, in contradistinction to what is in atomic physics, 
where it is $\approx \alpha_{em}.$  This numerical value is a cause for
concern if one is  to perform  the intended calculation by an 
approximate expansion either in $k/m_c$ or $m_c/k$.  We must calculate 
effects due to these potential terms exactly.  In this section, we 
shall show how this is handled in our formulation.

\subsection{Potential Terms and Energy Shifts for Single Insertions}

The expression for the energy shift due to a single insertion is

\begin{equation}
\Delta E=\int_{-\infty}^\infty d^3 \vec \xi \psi(\vec \xi) ^\dagger 
\tilde V(\xi)\psi(\vec{\xi}),
\end{equation}
with the Fourier transform
\begin{equation}
\tilde V(\xi)={1\over 2\pi^2}\int_{-\infty}^\infty d^3ke^{i\vec k \cdot \vec \xi}
V(k^2).
\label{eq4A.2}
\end{equation}
After introducing the momentum wave function as in Eq.(\ref{eq2.26}), 
\begin{equation}
\tilde \gamma_l(\xi)\xi ^{-l}=(\tilde \gamma_l(\xi)\xi ^{-l})^\star={1\over 
\sqrt{2\pi}}\int_{-\infty}^\infty dp \ e^{ip\xi}\gamma_l^l(p),
\end{equation}
into the spatial wave function
\begin{equation}
\psi(\vec \xi)=Y_{lm}(\Omega)\tilde \gamma_l(\xi)\xi^{-(l+1)},
\end{equation}
because the potential is spherically symmetrical, we can immediately
integrate over the solid angle $\Omega $ to obtain

\begin{equation}
\Delta E= {1\over 2\pi}\int _0^\infty d \xi \int_{-\infty}^\infty dp_1 e^{ip_1 \xi}
\gamma _l^l(p_1)\tilde V(\xi)\int_{-\infty}^\infty dp_2e^{ip_2 \xi}\gamma_l^l(p_2).
\end{equation}

At this point, we use the momentum representation of the potential in eq.(\ref{eq4A.2}).  
We write 
$\vec k \cdot \vec \xi=k\xi cos\theta$ and integrate over $\xi$ and the  
angles to obtain

\begin{eqnarray}
 \Delta E ={i\over 2\pi^2}
\int _{-\infty}^\infty dp_1 \gamma _l^l(p_1)
\int _{-\infty}^\infty dp_2 \gamma _l^l(p_2)
\int_o^\infty kdk \ ln({p_1+p_2+k \over p_1+p_2-k})V(k^2)
\end{eqnarray}
Let us  specialize and consider the energy shifts of the ground state, the wave
function of which is (n=1)

\begin{equation}
\gamma_{l=0}^{l=0}(p)=A_0{1\over (p-i\kappa_1)^2},
\label{eq4A.7}
\end{equation}
with
\begin{equation}
A_0=-\sqrt{2\kappa_1 ^3\over \pi}, \ \ \ 
\kappa_1={4\over 3}\alpha_s m_{reduced}={2\over 3}\alpha_s m_b 
\label{eq4A.8}
\end{equation}
for a bottomonium.
We can easily use the residue theorem to carry out the integrations. We shall write
as our standard form for any piece of the momentum space potential

\begin{equation}
V(k^2)={1\over k^2}g(k^2), 
\end{equation}
taking into account of 
the factor of $2\pi^2$ in Eq.(\ref{eq4A.2}), then it is easy to show that  the 
corresponding energy shift is 
\begin{equation}
{1\over 2\pi^2}< {1\over k^2}g(k^2)>=N\int_0^\infty dk{1\over( k^2+4\kappa_1^2)^2}
g(k^2),
\label{4A.10}
\end{equation}
where the  factor $N$ is
\begin{equation}
N=A_0^2 16\kappa_1. 
\end{equation}
\bigskip
The potential pieces to account for charm effects are due to:
(a) the charm loop in the lowest order

\begin{equation}
V_{charm}^{NLO}(k^2)=(-{4\over 3}\alpha_s)T_F\alpha_s {2\over 3\pi}
\big(\int_1^\infty dz {f(z)\over k^2+4m^2z^2}+{1\over k^2}( -{1\over 2}ln({k^2\over m^2})+{5\over 6})
\big),
\label{eq4A.12}
\end{equation}
where
\begin{equation}
f(z)={1\over z^2}\sqrt{z^2-1}(1+{1\over 2z^2}),
\end{equation}
(b) the iteration of the above and its combination with the zero mass quark 
and gluon effects \cite{Fishler}

\begin{equation}
V_{massless}^{NLO}(k^2)=(-{4\over 3}\alpha_s){\alpha_s\over 4\pi}{1\over k^2}
  \big( -\beta_0 ln({k^2\over \mu^2})+a_1 \big), 
\label{eq4A.14}
\end{equation}
where

\begin{equation}
\beta_0={11\over3}C_A -{4\over 3}T_F n_l, \ \ a_1={31\over 9}C_A-{20\over 9}T_F n_l,
\nonumber
\end{equation}
and $\mu$ is a subtraction scale, $C_A=3, T_F=1/2$ and $n_l=4$, and 
(c) $\alpha_s$ corrections to the charm loop.  In this section, the mass symbol
m refers to the mass of the charm quark.  We have \cite{Melles, Hoang3}

\begin{equation}
V_{charm}^{NNLO}=V_{charm}^{NNLO}(1)+V_{charm}^{NNLO}(2)
+V_{charm}^{NNLO}(3),
\end{equation}
where
\begin{eqnarray}
V_{charm}^{NNLO}(1) = \bar c_1 &[& \bar a_1
            \int_1^\infty dz f(z) {ln({k^2\over 4m^2})\over k^2+4m^2z^2}
            + \bar a_2\int _1^\infty dz{f(z)\over k^2+4m^2z^2} \nonumber \\
 & &
     + \bar a_3{1\over k^2}ln^2({k^2\over m^2})
            + \bar a_4 {1\over k^2}ln({k^2\over m^2}) + \bar a_5{1\over k^2}],
\end{eqnarray}
with
$$\bar c_1=(-4\alpha_s/3)({\alpha_s\over 4 \pi})(T_F\alpha_s {2\over 3 \pi}) ,
$$
$$
\bar a_1=  - 2\beta_0,\ \ \bar a_2=  - 2\beta_0 ln({4m^2\over \mu^2}) + 2a_1, \ \ 
  \bar a_3=\beta_0 ,
$$
  $$\bar a_4=- (5/3)\beta_0 + \beta_0 ln({m^2\over \mu^2}) - a_1, \ \ 
  \bar a_5=- (5/3)\beta_0 ln({m^2\over \mu^2}) + (5/3)a_1 ,$$

\begin{eqnarray}
 V_{charm}^{NNLO}(2) &=& \bar c_2[\int_1^\infty dz_1 dz_2{k^2f(z_1)\over k^2+4m^2z_1^2}
     {f(z_2)\over k^2+4m^2z_2^2}
               -\int_1^\infty dz f(z){ln(k^2/4m^2)\over k^2+4m^2z^2}\nonumber \\
 & &
                  +(-2ln2+5/3)\int_1^\infty dz {f(z)\over k^2+4m^2z^2}
              +{1\over 4}{1\over k^2}(-ln(k^2/m^2)+5/3)^2],
\end{eqnarray}
where
\begin{equation}
\bar c_2=(-{4\over 3}\alpha_s)(T_F\alpha_s{2\over 3 \pi })^2, \nonumber
\end{equation}
and
\begin{equation}
V_{charm}^{NNLO}(3)=\bar c_3{1\over k^2}[c_1ln(1+{k^2\over 4c_2^2m^2})
+d_1ln(1+{k^2\over 4d_2^2m^2})
-\big( ln({k^2\over m^2})-{161\over 114}-{26\over 19}\zeta_3\big)],
\end{equation}
where 
\begin{equation}
\bar c_3=(-{4\over 3}\alpha_s)({\alpha_s\over 4\pi})^2
({76\over 3}T_F),\ \ c_1={ln{A\over d_2}\over ln{c_2\over d_2}}, \ \ 
d_1={ln{c_2\over A}\over ln{c_2\over d_2}}, \nonumber
\end{equation}
and
\begin{equation}
A=exp({161\over 228}+{13\over 19}\zeta_3-ln2), 
\ \ c_2=0.470\pm 0.005, \ \ d_2=1.120\pm 0.010. \nonumber
\end{equation}

\smallskip
We now apply Eq.(\ref{4A.10}) to obtain energy shifts.  The more tedious task in the 
calculation is to integrate over the spectral density $f(z)$.  It turns out that 
all the integrals involved can be evaluated analytically into simple functions.
Our method is first to assume that the parameter 

\begin{equation}
y\equiv {\kappa_1\over m}
\end{equation}
is $\le 1$, so that we can expand the integrand in powers of $y/z$.  The
integration over z can then be performed and the infinite series are then
resummed.  Since we have combinations of integrals which correspond to 
physical quantities and the analytic results for them, 
we can then unambiguously continue them into values of $y\ge 1$, which are relevant for 
investigating the toponium and the very light quark mass limit.  

For the next to leading order, we have 

\begin{equation}
(\Delta E)_{charm}^{NLO} 
 =(-{4\over 3}\alpha_s)T_F \alpha_s
{2 \over 3\pi}\kappa_1\big(y^2I-(ln(2y)-{11\over 6})\big), 
\end{equation}
where \cite{Eiras2}

\begin{eqnarray}
I &=&-{1\over y^2}({2\over y^2}+{11\over 6})-
{1\over (1-y^2)^{1/2}}{cos^{-1}y\over y}
({2\over y^4}+{1\over 2y^2}-1)\nonumber \\
 & &
+{\pi\over 2}{1\over y^3}({2\over y^2}+{3\over2})
, \ \ \ \ \ \ \  y\le 1
\end{eqnarray}
which goes to $2/5$ as $y \to 0$, reproducing the well-known QED Uehling result \cite{Uehling}.  
By analytic continuation, we have 

\begin{eqnarray}
I &=&-{1\over y^2}({2\over y^2}+{11\over 6})-
{1\over (y^2-1)^{1/2}}{ln(y+(y^2-1)^{1/2})\over y}
({2\over y^4}+{1\over 2y^2}-1) \nonumber \\
& &
+{\pi\over 2}{1\over y^3}({2\over y^2}+{3\over2})
, \ \ \ \ \ \ \  y\ge 1
\end{eqnarray}
which will be useful when we consider $t \bar t $ threshold or high 
Z muonic atoms.  The result here agrees with what was obtained before
\cite{Hoang3}.

For the next-next leading order, we just list the analytic result for each
piece.  They are

\begin{eqnarray}
{1\over 2\pi^2}<\int_1^\infty dz f(z) {ln(k^2/4m^2)\over k^2+4m^2z^2}>&=& N
\int_1^\infty dz f(z) \int _0^\infty dk
{k^2\over (k^2+4\kappa_1^2)^2}{ln(k^2/4m^2)\over k^2+4m^2z^2}\nonumber \\
 &=& N{2\over( 2m)^3}\int_1^\infty dz f(z) \int _0^\infty dk'
{k'^2\over (k'^2+y^2)^2}{ln(k')\over k'^2+z^2} \nonumber \\
 &=& N{2\over( 2m)^3}\int_1^\infty dz f(z)
{\pi\over 4y}[{z^2+y^2\over (z^2-y^2)^2}ln(y)
\nonumber \\ & & ~~~~~~~~~~
\ -{2yz\over (z^2-y^2)^2}ln(z)+{1\over z^2-y^2}]\nonumber \\
 &=& N{2\over( 2m)^3}{\pi\over 4y}[-2y{d\over dy^2}H
+(ln(y)(2y^2{d\over dy^2}+1)+1)G], \nonumber \\
\end{eqnarray}

\begin{eqnarray}
{1\over 2\pi^2}<\int_1^\infty dz f(z) {1\over k^2+4m^2z^2}>&=& 
N\int_1^\infty dz f(z) \int _0^\infty dk
{k^2\over (k^2+4\kappa_1^2)^2}{1\over k^2+4m^2z^2}\cr &
=&N{1\over( 2m)^3}({\pi\over 2y})\int_1^\infty dz f(z)[{y^2- yz\over (z^2-y^2)^2}]
+{1\over 2}{1\over z^2-y^2}\cr &
=&N{1\over( 2m)^3}(-{\pi\over 4y}){d\over dy}(J-yG),
\end{eqnarray}

\begin{eqnarray}
{1\over 2\pi^2}<{1\over k^2}ln^2({k^2\over m^2})>&
=&N\int _0^\infty dk {ln^2({k^2\over m^2})\over (k^2+4\kappa_1^2)^2}\nonumber \\ &
=&N{\pi \over 4\kappa_1^3}[{\pi^2\over 8}+{1\over 2}ln^2(2y)-ln(2y)],
\end{eqnarray}

\begin{eqnarray} {1\over 2\pi^2}<{1\over k^2}ln({k^2\over m^2})>&
=&N\int _0^\infty dk {ln({k^2\over m^2})\over (k^2+4\kappa_1^2)^2}\nonumber \\ &
=&N{\pi \over 16\kappa_1^3}[ln(2y)-1],
\end{eqnarray}

\begin{eqnarray} 
{1\over 2\pi^2}<{1\over k^2}ln(1+{k^2\over 4m^2d^2})>&
=& N\int _0^\infty dk {ln(1+{k^2\over 4m^2d^2})\over (k^2+4\kappa_1^2)^2}\nonumber \\ &
=& N{\pi \over 16 m^3 y^2}[{1\over y}ln(1+{y\over d})-{1\over y+d}],
\end{eqnarray}

\begin{equation}
{1\over 2\pi^2}<{1\over k^2}>=N{\pi \over 32\kappa_1^3},
\end{equation}
and

\begin{eqnarray} 
& &{1\over 2\pi^2} <\int_1^\infty dz_1 \int_1^\infty dz_2 f(z_1)f(z_2){k^2\over (k^2+4m^2z_1^2)
(k^2+4m^2z_2^2)}>\nonumber \\ &
=&<\int_1^\infty dz_1 \int_1^\infty dz_2 f(z_1)f(z_2)({1\over z_1^2-z_2^2})
({z_1^2\over k^2+4m^2z_1^2}-{z_2^2\over k^2+4m^2z_2^2})>\nonumber \\ &
=&N{\pi\over 16m^3y}\int_1^\infty dz_1 \int_1^\infty dz_2 f(z_1)f(z_2)
[(y^2{d\over dy^2}+{1\over 2})(-y^2{1\over z_1^2-y^2}{1\over z_2^2-y^2})\nonumber \\ &
& ~~~~~~~~~~~~~~~~~~~~~~
\ -y{d\over dy^2}({1\over z_1+z_2}{z_2^2\over z_2^2-y^2}
-y^2{z_1\over z_1^2-y^2}{1\over z_2^2-y^2})] \nonumber \\ &
=&N{\pi\over 32m^3y}[(y{d\over dy}+1)(-y^2G^2)-{d\over dy}
(K-y^2GJ)].
\end{eqnarray}

\bigskip
The functions G, H, J, K which individually can be defined only for $y\le1$ are 

\begin{eqnarray} 
G& \equiv&\int _1^\infty dz {f(z)\over z^2-y^2}\nonumber \\ &
=&{1\over y^2}[(1+{1\over 2y^2})(1-{(1-y^2)^{1/2}\over y}sin^{-1}(y))-{1\over 6}],
\end{eqnarray}

\begin{eqnarray} 
H &\equiv& \int _1^\infty dz{zf(z)ln(z)\over z^2-y^2}\nonumber \\ &
=&-{\pi\over 8y^2}[4(1-y^2)^{1/2}ln(1+(1-y^2)^{1/2})
-3ln(2)+{1\over 2}\nonumber \\ &&\ \ \ \ +{2\over y^2}((1-y^2)^{1/2}ln(1+(1-y^2)^{1/2})
-ln(2))],
\end{eqnarray}

\begin{eqnarray}
J&\equiv&\int_1^\infty dz{zf(z)\over z^2-y^2}
\nonumber \\ &=&{\pi\over 2}{1\over y^2}[{3\over 4}-(1-y^2)^{1/2}
+{1\over 2y^2}(1-(1-y^2)^{1/2})].
\end{eqnarray}

\begin{eqnarray} 
K& \equiv&\int_1^\infty dz_1 dz_2 f(z_1)f(z_2){1\over z_1+z_2}
{z_2^2\over z_2^2-y^2}\nonumber \\ &
=&\pi[ ( {1\over 8y^6} + {19\over 48y^4} + {9\over 80y^2} )
 +sin^{-1}(y)
   ( -{ 1\over 8y^7} - {3\over 8y^5} + {1\over 2y} )\nonumber \\ &&
 \ \ \ \ +sin^{-1}(y)(1-y^2)^{1/2}
  ( - {1\over 8y^7} - {7\over 16y^5} - {3\over 8y^3} )\nonumber \\ &&
\ \ \ \  +(1-y^2)^{1/2}
  ( {1\over 8y^6} + {11\over 24y^4} + {5\over 12y^2} )]. 
\end{eqnarray}

Using them, we have the following results for $y\le 1$:

\begin{eqnarray} 
& & -2y{d\over dy^2}H
+(ln(y)(2y^2{d\over dy^2}+1)+1)G \nonumber \\
&= &
   {1\over 2y^4} - {2\over y^4}ln(y) + {5\over 6y^2} - {11\over 6y^2}ln(y)
 + \pi ( {ln(2)\over y^5} + {3ln(2)\over 4y^3} - {1\over 4y^5} - {5\over 8y^3} )\nonumber \\ & &
 + g_2(1-y^2)^{1/2}
  ( {2\over y^5} + {2\over y^3} )
 + g_2(1-y^2)^{-1/2}
   ( {1\over 2y^3} + {1\over y} )
 + g_1(1-y^2)^{1/2}
   ( - {1\over 2y^5} -{1\over  y^3 }),
\end{eqnarray}
 
\begin{eqnarray} 
{d\over dy}(J-yG) &=& 
   {2\over y^4} + {11\over 6y^2}
 + \pi( - {1\over y^5} - {3\over 4y^3 })
 + g_1(1-y^2)^{1/2}
  ( - {2\over y^5} - {2\over y^3}) \nonumber \\ & &
 + g_1(1-y^2)^{-1/2}
  ( - {1\over 2y^3} - {1\over y} ),
\end{eqnarray}

\begin{eqnarray} 
& &(y{d\over dy}+1)(-y^2G^2)- {d \over dy}
(K-y^2GJ)\nonumber \\
   &=&
          { 7\over 4y^6} + {13\over 3y^4} + {85\over 36y^2} 
         +\pi
          ( -{ 2\over 5y^3})
+ \pi^2( - {7\over 16y^8} - {15\over 16y^6} + {1\over 4y^2} )\nonumber \\ & &
+ (1-y^2)^{-1/2}g_1
          ( - {7\over 2y^7} - {19\over 3y^5} + {13\over 6y^3} + {11\over 3y} )
+ g_1^2
         ( {7\over 4y^8} + {15\over 4y^6} - {1\over y^2} ), 
\end{eqnarray}
where 
\begin{equation}
g_1=sin^{-1}(y)-{\pi\over 2}=-cos^{-1}(y),
\end{equation}
and 
\begin{eqnarray}
g_2 &=& ln(y) \ sin^{-1}(y)-{\pi\over 2} \ ln(1+(1-y^2)^{1/2})\nonumber \\
&=& -ln(y)cos^{-1}(y)-{\pi \over 2}ln({1\over y}+(({1\over y})^2-1)^{1/2}).
\end{eqnarray}
This completes our evaluation of the expectation values of various pieces 
of potentials into simple functions.  Please note that  $g_1, g_2$ and
$(1-y^2)^{1/2}$ appear together in the correct combinations to allow us 
to analytically continue the energy shifts into real functions on the proper 
branch for $y\ge1$. They are given by the substitutions
\begin{equation}
{cos^{-1}(y)\over (1-y^2)^{1/2}} \ \  \ y\le 1 \ \ \ \ \rightarrow \ \ \ 
{ln(y+(y^2-1)^{1/2}) \over (y^2-1)^{1/2}} \ \ \ y\ge 1,
\end{equation} 
and 
\begin{equation}
{ln({1\over y}+(({1\over y})^2-1)^{1/2}) \over (({1\over y})^2-1)^{1/2}}
\ \  \ y\le 1 \ \ \ \ \rightarrow \ \ \ 
{cos^{-1}({1\over y})\over (1-({1\over y})^2)^{1/2}} \ \  \ y\ge 1.
\end{equation}

We have a complete analytical result for the single insertion of the potential to 
account for the finite charm mass effects for all values of y.  Some of the integrals 
were not given in analytical forms by other authors and can be calculated numerically [6].  
The values agree with ours within numerical accuracy, although we are somewhat unsure
how the numerical 
integration programs decide on what branch to follow for $y\ge 1$.

\subsection{Potential Terms and Energy Shifts for Double Insertions}

In the next-next leading order, we have energy shifts due to double insertions
of $V^{NLO}_{charm}$ and $V^{NLO}_{massless}$ of Eqs.(\ref{eq4A.12}) and (\ref{eq4A.14}), 
respectively.
For a state with wave function $\psi_{E_{n=1}}$ and perturbing potential $\tilde V$, 
the shift is given by 
\begin{equation}
\Delta E_{double \ insertion}=-\int d^3\vec x \ d^3\vec y\ 
\psi_{E_{n=1}}^\dagger (\vec x)\tilde V(x)
Lim_{E\to E_{n=1}}G(\vec x, \vec y; E)\tilde V(y)\psi_{E_{n=1}}(\vec y).
\end{equation}
Note that the regularization applies to the Green's function only.  We shall discuss this
point later on when we compare results.
We write 
\begin{eqnarray} 
\tilde V(x)&=&{1\over 2\pi^2}\int d^3\vec k_1 e^{i\vec k_1 \cdot \vec x} V(k_1^2),
\nonumber \\ & =&({1\over 2\pi^2}){2\pi \over ix}\int_0^\infty  dk_1 k_1(e^{ikx}-e^{-ikx}) V(k_1^2)
\nonumber \\ &=&({1\over 2\pi^2}){2\pi \over ix}\int_{-\infty} ^\infty  dk_1 k_1e^{ikx} V(k_1^2)
\end{eqnarray}
and a similar expression for $\tilde V(y)$ with  $k_1\to k_2$.  Likewise, we express
\begin{equation}
\psi (\vec x)^\dagger=Y_{l m}^\dagger (\Omega_x)({1\over x})({1\over 2\pi})^{1/2}
\int_{-\infty}^\infty dpe^{ipx}\gamma_l^l(p),
\end{equation}
and for $\psi(\vec y)$ with $p \to p'$.  Upon using the addition 
theorem 
\begin{equation}
P_l({\vec x \cdot \vec y\over x y})={4\pi \over 2l+1}
\sum _mY_{l m}^\dagger (\Omega_x)Y_{l m} (\Omega_y),
\end{equation}
to carry out the angular integration of $\vec x$ and $\vec y$, we arrive at
\begin{eqnarray}
\Delta E_{double \ insertion}&=&-{8\pi^2\over (2\pi^2i)^2}
\int_0^\infty dx\int_0^\infty dy\int_{-\infty}^\infty dk_1 k_1
\int_{-\infty}^\infty dk_2 k_2
\int_{-\infty}^\infty dp\int_{-\infty}^\infty dp'\nonumber \\ 
& &\times e^{ik_1x}
e^{ik_2y}e^{ipx}e^{ip'y}\gamma_l^l(p)\gamma_l^l(p') V(k_1^2)
 V(k_2^2)G_l(x,y;E).
\end{eqnarray}

Now we specialize to the ground state given by Eqs.(\ref{eq4A.7}) and 
(\ref{eq4A.8}).
We integrate over x and y, by using 
\begin{equation}
\int_0^\infty dx e^{i(k_1+p+p_1)x}=i{1\over k_1+p+p_1},
\end{equation}
and Eq.(\ref{eq3.8}), which give 
\begin{eqnarray}
\Delta E_{double \ insertion}&=&{2m_b\over (2\pi^2)^2}{\kappa_1^3\over \kappa _0}
({1\over 2\pi i})^2
\int_0^1 d\rho \rho^{-1-\nu}\nonumber \\ & & \times
\int_{-\infty}^\infty dk_1 k_1
\int_{-\infty}^\infty dk_2 k_2
\int_{-\infty}^\infty dp\int_{-\infty}^\infty dp'
\int_{-\infty}^\infty dp_1\int_{-\infty}^\infty dp_2
\nonumber \\ & &\times ({1\over p-i\kappa_1})^2({1\over p'-i\kappa_1})^2
{1\over k_1+p+p_1}{1\over k_2+p'+p_2}
ln({bc\over ad})
 V(k_1^2) V(k_2^2)\nonumber \\ &
=&{2m_b\over (2\pi^2)^2}{\kappa_1^3\over \kappa _0}
\int_0^1 d\rho \rho^{-1-\nu} 
\int_{-\infty}^\infty dk_1 k_1
\int_{-\infty}^\infty dk_2 k_2
\int_{-\infty}^\infty dp_1\int_{-\infty}^\infty dp_2
\nonumber \\ & &\times 
({1\over k_1+p_1+i\kappa_1})^2({1\over k_2+p_2+i\kappa_1})^2 ln({bc\over ad})
 V(k_1^2) V(k_2^2).
\label{eq4B.7}
\end{eqnarray}

To organize our calculation better in what follows and so as not to have to track the branch cuts 
of products of logarithmic functions, we shall write
\begin{eqnarray} 
V_{charm}^{NLO}(k^2)&=&(-{4\over 3}\alpha_s)T_F\alpha_s {2\over 3\pi}Lim_{\mu_i\to 0}
\big(\int_1^\infty dz {f(z)\over k^2+4m^2z^2}-{1\over 2}
\int _0^1{dt\over t}{1\over k^2+\mu_i^2/t}\nonumber \\ 
& &~~~~~~~~~~~~~~~~~~~~~~~~~~~~~~~~~~~~~
+{1\over k^2}( {1\over 2}ln({m^2\over \mu_i^2})+{5\over 6})
\big),
\label{eq4B.8}
\end{eqnarray}
and

\begin{equation}
V_{massless}^{NLO}(k^2)=(-{4\over 3}\alpha_s){\alpha_s\over 4\pi}
Lim_{\mu_i \to 0}\big(-\beta_0\int_0^1{dt\over t}
{1\over k^2+\mu_i^2/t}+{1\over k^2}
  ( -\beta_0 ln({\mu_i^2\over \mu^2})+a_1) \big). 
\end{equation}

Therefore, it is clear that after we calculate the energy shift due 
to
\begin{equation} 
V_1(k^2)=(-{4\over 3}\alpha_s)(\alpha_s T_F {2\over 3\pi})
\int_1^\infty dzf(z){1\over k^2+\bar z^2}, \ \ \  \bar z=2mz,
\end{equation}
we can  make minor changes on the spectral density and take $\bar z ^2 \to \mu_i^2/t$
or $\bar z\to 0$ to get the other contributions.  This is what we are going to
do immediately.

The integration over $k_1$ in Eq.(\ref{eq4B.7}) gives
\begin{equation}
\int_{-\infty}^\infty dk_1 k_1{1\over k_1^2+\bar z_1^2}
({1\over k_1+p_1+i\kappa_1})^2=i\pi({1\over p_1+i(\kappa_1+\bar z_1)})^2,
\end{equation}
and a similar result for  $k_2$.  This leads to integrations over $p_1$ and $p_2$,
which give a factor of 
\begin{equation}
(2\pi i)^2{d^2\over dp_1 dp_2}ln({bc\over ad}),
\end{equation}
evaluated at $p_1=-i(\kappa_1+\bar z_1)$ and $p_2=-i(\kappa_1+\bar z_2)$.  

\bigskip
We are interested in the limit of 
\begin{equation}
p_0=Lim_{\epsilon \to 0}i\kappa_1(1+\epsilon), \ \ \epsilon=1-\nu.
\end{equation}
Thus in the calculation of the energy shift, we shall keep only the $\epsilon ^0$ 
and the ${1\over \epsilon}$ terms.  Since the $\rho$ integration will 
produce the $1\over \epsilon$ pole, we must keep terms to order $\epsilon ^1$ 
in the integrand as well.  Now it is also clear that all terms of order  
 $\epsilon \rho^n $ with $n\ge 2 $  can be dropped. 
We shall show later on that 
the ${1\over \epsilon } $ terms cancel, because we are interested in 
the subtracted propagator
\begin{equation}
Lim_{E\to E_{n=1}}(-G(\vec x,\vec y;E)-|E_{n=1}>{1\over E- E_{n=1}}<E_{n=1}|),
\label{eq4B.14}
\end{equation}
to obtain the regulated level shift.

Carrying out the differentiation and evaluating at the value of $p_{1,2}$ as 
discussed, we find 
\begin{eqnarray}
{d^2\over dp_1 dp_2}ln({bc\over ad})&=&\rho  \left [
-{4\kappa_1^2\over ((2\kappa_1+\bar z_1)(2\kappa_1+\bar z_2)-\rho \bar z_1\bar z_2)^2}
-\epsilon  {8\kappa_1^2\over (2\kappa_1+\bar z_1)^2(2\kappa_1+\bar z_2)^2} 
\right .
\nonumber \\ 
& & \left.
+\epsilon {8\kappa_1^3(2\kappa_1+\bar z_1+2\kappa_1+\bar z_2)\over 
(2\kappa_1+\bar z_1)^3(2\kappa_1+\bar z_2)^3} \right].
\end{eqnarray}

We are now faced with an integral
\begin{eqnarray}
I_\rho\equiv-4\kappa_1^2\int_0^1d\rho \rho^{-\nu} & &\left[
{1\over ((2\kappa_1+\bar z_1)(2\kappa_1+\bar z_2)-\rho \bar z_1\bar z_2)^2}
+\epsilon  {2\over (2\kappa_1+\bar z_1)^2(2\kappa_1+\bar z_2)^2} \right.
\nonumber \\
& & \left. -\epsilon {2\kappa_1(2\kappa_1+\bar z_1+2\kappa_1+\bar z_2)\over 
(2\kappa_1+\bar z_1)^3(2\kappa_1+\bar z_2)^3} \right],
\end{eqnarray}
which can be easily calculated by adding and subtracting the terms in the square brackets at $\rho =0.$  Putting these pieces 
together, we obtain
\begin{eqnarray}
\Delta E(V_1 V_1)&=& 2m_b{\kappa_1^3\over \kappa _0}
\left((      {4\over 3}\alpha_s)(\alpha_s T_F {2\over 3\pi})\right)^2
\nonumber \\
& & \times \int_1^\infty  dz_1f(z_1)\int_1^\infty  dz_2 f(z_2)4\kappa_1^2
\left\{
{1\over (2\kappa_1+\bar  z_1)^2(2\kappa_1+\bar z_2)^2} \right. 
\nonumber \\
& & \times \left[-{1\over \epsilon}
+Ln ({(2\kappa_1+\bar z_1)(2\kappa_1+\bar z_2)-\bar z_1 
\bar z_2\over (2\kappa_1+\bar z_1)(2\kappa_1+\bar z_2)})
\right.
\nonumber \\
& & \left. -{\bar z_1 \bar z_2\over (2\kappa_1+\bar z_1)(2\kappa_1
+\bar z_2)-\bar z_1\bar z_2} \right]
-{2\over (2\kappa_1+\bar z_1)^2(2\kappa_1+\bar z_2)^2}
\nonumber \\
& & \left.+ 2\kappa_1\left({1\over  
(2\kappa_1+\bar z_1)^2(2\kappa_1+\bar z_2)^3}+
{1\over  
(2\kappa_1+\bar z_1)^3(2\kappa_1+\bar z_2)^2}\right) \right\}.
\label{eq4B.17}
\end{eqnarray}

There remain two items to be settled.  We first expand
\begin{equation}
\kappa_0^{-1} = \kappa_1^{-1}(1-\epsilon)
\end{equation}
in Eq.(\ref{eq4B.17}) and then perform a subtraction as indicated in 
Eq.(\ref{eq4B.14}) because the ground state should 
not be included in the propagator. This yields the 
subtraction term
\begin{eqnarray} 
& & -\int d^3\vec x\int d^3\vec y 
\psi^\dagger (\vec x)_{ E_{n=1}}\tilde V_1(x)\psi (\vec x)_ {E_{n=1}} 
{1\over E-E_{n=1}}\psi^\dagger (\vec y)_ {E_{n=1}}\tilde V_1(y)\psi (\vec y)_ {E_{n=1}}
\nonumber \\
&=&{m_b\over 2\kappa_1^2}({1\over \epsilon})(1-{3\over 2}\epsilon )
\left(\Delta E^{NLO}\right)^2,
\end{eqnarray}
where
\begin{equation}
(\Delta E)^{NLO}=(-{4\over 3}\alpha_s)(T_F \alpha_s {2\over 3\pi})(4\kappa_1^3)
\int_1^\infty  dz f(z){1\over (2\kappa_1+\bar z)^2}.
\end{equation}
The end result is
\begin{eqnarray}
\Delta E(V_1 V_1)^{reg} &=&
2m_b\kappa_1^2
(({4\over 3}\alpha_s)(\alpha_s T_F {2\over 3\pi}))^2
\nonumber \\
& &\times \int_1^\infty  dz_1f(z_1)\int_1^\infty  dz_2 f(z_2)4\kappa_1^2
\left\{
{1\over (2\kappa_1+\bar  z_1)^2(2\kappa_1+\bar z_2)^2} \right.
\nonumber \\
& &\times \left[-{1\over 2}
+Ln ({(2\kappa_1+\bar z_1)(2\kappa_1+\bar z_2)-\bar z_1 
\bar z_2\over (2\kappa_1+\bar z_1)(2\kappa_1+\bar z_2)}) \right.
\nonumber \\
& & \left. -{\bar z_1 \bar z_2\over (2\kappa_1+\bar z_1)(2\kappa_1
+\bar z_2)-\bar z_1\bar z_2} \right]
-{2\over (2\kappa_1+\bar z_1)^2(2\kappa_1+\bar z_2)^2}
\nonumber \\
& & \left.+ 2\kappa_1\big({1\over  
(2\kappa_1+\bar z_1)^2(2\kappa_1+\bar z_2)^3}+
{1\over  
(2\kappa_1+\bar z_1)^3(2\kappa_1+\bar z_2)^2}\big) \right\},
\label{eq4B.21}
\end{eqnarray}
As a check that our regularization is correct, we
look into the modified Coulomb potential 
\begin{equation}
\tilde V(\vec x)=-{4\over 3} \alpha_s(1+\delta){1\over x},
\end{equation}
with the term proportional to $\delta$ treated as a perturbation, 
which produces the ground energy
\begin{equation}
E_{ground}=-{m_b\over 4}({4\over 3}\alpha_s)^2(1+\delta)^2.
\end{equation}
The $\delta^2$ term is recovered from Eq.(\ref{eq4B.21}) by setting $\bar z_{1,2}\to 0$
and requiring the spectral density to satisfy $(\alpha_s T_F {2\over 3\pi})\int dz f(z)\to1$.  

\smallskip
We shall take this as our standard subtraction, 
which will be applied to 
Eq.(\ref{eq4B.17}) and remaining energy shifts, by the operation
\begin{equation}
\kappa_0 \ \ \ \to \ \ \ \kappa_1,  \ \ \ \ \   {1\over \epsilon} \ \ \  \to \ \ \ 
{1\over 2}.
\label{eq4B.24}
\end{equation} 

To finish the calculation due to Eq.(\ref{eq4B.8}), we define
\begin{equation}
V_{charm}^{NLO}(k^2)=V_1(k^2)+V_2(k^2)+V_3(k^2),
\end{equation}
\begin{equation}
V_2(k^2)=(-{4\over 3}\alpha_s)T_F\alpha_s {2\over 3\pi}Lim_{\mu_i\to 0}
\big(-{1\over 2}
\int _0^1{dt\over t}{1\over k^2+\mu_i^2/t}\big),
\end{equation}
\begin{equation}
V_3(k^2)=(-{4\over 3}\alpha_s)T_F\alpha_s {2\over 3\pi}Lim_{\mu_i\to 0}
{1\over k^2}( {1\over 2}ln({m^2\over \mu_i^2})+{5\over 6}),
\end{equation}
and 
\begin{equation}
bb=2m_b{\kappa _1^3\over \kappa_0}({4\over 3}\alpha_s(\alpha_s T_F {2\over 3\pi}))^2,
\end{equation}
then the shift due to $V_1V_2$ is
\begin{eqnarray}
\Delta E(2V_1 V_2)&=&2(-1/2)bb
 \int_1^\infty  dz_1f(z_1)\int_0^1{dt\over t}4\kappa_1^2 \left\{
{1\over (2\kappa_1+\bar  z_1)^2(2\kappa_1+\mu_i/\sqrt t)^2} \right.
\nonumber \\
& & \times \left[-{1\over \epsilon}
+Ln ({(2\kappa_1+\bar z_1)(2\kappa_1+\mu_i/\sqrt t)-\bar z_1 
\mu_i/\sqrt t\over (2\kappa_1+\bar z_1)(2\kappa_1+\mu_i/\sqrt t)})
\right.
\nonumber \\
& & \left. \ \ \ \ -{\bar z_1 \mu_i/\sqrt t\over (2\kappa_1+\bar z_1)(2\kappa_1
+\mu_i/\sqrt t)-\bar z_1\mu_i/\sqrt t} \right]-{2\over 
(2\kappa_1+\bar z_1)^2(2\kappa_1+\mu_i/\sqrt t)^2}
\nonumber \\
& & \left. +2\kappa_1\left( {1\over(2\kappa_1+\bar z_1)^3
(2\kappa_1+\mu_i/\sqrt t)^2}
+{1\over(2\kappa_1+\bar z_1)^2(2\kappa_1+\mu_i/\sqrt t)^3}\right) \right\} .
\end{eqnarray}

We perform the t integration to obtain
\begin{eqnarray}
\Delta E(2V_1 V_2) &=& bb
 \int_1^\infty  dzf(z) \left\{ {1\over (2\kappa_1+\bar  z)^2}
\left[2Li({\bar z\over 2\kappa_1+\bar z})-2({1\over 2}+
ln({\mu_i\over 2\kappa_1}))
\right. \right.
\nonumber \\
& & \left.\left. -{2\over \epsilon}(1+ln({\mu_i\over 2\kappa_1})) \right] 
+{4\kappa_1\over (2\kappa_1+\bar z)^3}
(1+ln({\mu_i\over 2\kappa_1}))
\right\}.
\label{eq4B.30}
\end{eqnarray}
Similarly, the $V_1V_3$ shift is arrived at by taking the 
zero mass limit of $\bar z_2$, which leads to
\begin{eqnarray}
\Delta E(2V_1 V_3)=  bb
 \int_1^\infty  dz f(z)[{1\over (2\kappa_1+\bar  z)^2}
(-{2\over \epsilon}-2)+{4\kappa_1\over (2\kappa_1+\bar  z)^3} ]
({1\over 2}ln({m^2\over \mu_i^2})+{5\over 6}))].
\label{eq4B.31}
\end{eqnarray}
It is important to note that $ln(\mu_i)$'s cancel in 
Eqs.(\ref{eq4B.30}-\ref{eq4B.31}), as they must,
because $\mu_i$ is introduced to facilitate our calculation.  Physical results
should not depend on it. Shifts due to other pieces are deduced in a similar way
upon being careful not to interchange the $\mu_i \to 0$ limit and the t-integration.  
We have
\begin{eqnarray}
\Delta E(V_2V_2+2V_2 V_3+V_3V_3) & = & bb{1\over 4\kappa_1^2} 
\left[-{1\over \epsilon} \left({11\over 6}+ln({m\over 2\kappa_1}) \right)^2
+ln({m\over 2\kappa_1})+{11\over 6} \right.
\nonumber \\
& & \left. - \left(\zeta_3-{\pi^2\over 6}+1\right)\right],
\end{eqnarray}
and
\begin{eqnarray}
\Delta E(2V_1( V_2+V_3)) &= &bb
 \int_1^\infty  dz f(z) \left[  {1\over (2\kappa_1+\bar  z)^2}
\left(2Li({\bar z\over 2\kappa_1+\bar z}) -
{2\over \epsilon}  \left({11\over 6}+ln({m\over 2\kappa_1})\right) 
\right .\right.
\nonumber \\
 & & \left.\left.
-2 ln({m\over 2\kappa_1})-{8\over 3} \right)
+{4\kappa_1\over (2\kappa_1+\bar  z)^3} 
 \left(ln({m\over 2\kappa_1})+{11\over 6}\right) \right].
\end{eqnarray}

In a similar manner, we introduce 
\begin{equation}
V_{massless}^{NLO}(k^2)=V_1'(k^2)+V_2'(k^2),
\end{equation}
where 
\begin{equation}
V_1'(k^2)=(-{4\over 3}\alpha_s){\alpha_s\over 4\pi}Lim_{\mu_i\to 0}
\big(-\beta_0 \int_0^1 {dt\over t}{1\over k^2+\mu_i^2/t}\big),
\end{equation}
and
 \begin{equation}
V_2'(k^2)=(-{4\over 3}\alpha_s){\alpha_s\over 4\pi}Lim_{\mu_i\to 0}
\big({1\over k^2}(-\beta_0ln({\mu_i^2\over \mu^2})+a_1) \big),
\end{equation}
and also
\begin{equation}
ab=2m_b{\kappa_1^3\over \kappa_0}({4\over 3}\alpha_s)^2(T_F\alpha_s{2\over 3\pi})
({\alpha_s\over 4\pi}). 
\end{equation}

We have 
\begin{eqnarray}
\Delta E(2V_1(V_1'+V_2')) &=& 2(ab)\int_1^\infty dzf(z) \left[ 
 {1\over (2\kappa_1+\bar z)^2}
\left( 2\beta_0 Li({\bar z\over 2\kappa_1+\bar z}) \right. \right.
\nonumber \\
& &\left.  -{1\over \epsilon}
\left(2\beta_0+a_1+2\beta_0 ln({\mu\over 2\kappa_1})\right)
 -2\beta_0ln({\mu \over 2\kappa_1})-\beta_0-a_1 \right)
\nonumber \\
& & \left. +{2\kappa_1\over (2\kappa_1+\bar z)^3}
 \left(2\beta_0 ln({\mu\over 2\kappa_1})+2\beta_0+a_1 \right) \right],
\end{eqnarray} 
and
\begin{eqnarray}
\Delta E(2(V_2+V_3)(V_1'+V_2')) &=& 2(ab){1\over 4\kappa_1^2}
\left[ -{1\over \epsilon} \left({11\over 6}+ln({m\over 2\kappa_1})\right)
\left(2\beta_0+a_1+2\beta_0ln({\mu \over 2\kappa_1})\right)
\right.
\nonumber \\
& & \left.
+\beta_0 ln ({m \mu \over (2\kappa_1)^2})+{17  \over 6}\beta_0+{a_1\over 2}
 -2\beta_0 \left(\zeta_3-{\pi^2\over 6}+1\right) \right].
\label{eq4B.39}
\end{eqnarray}
 
We see again that $\ln(\mu_i)$'s cancel.   The subtracted results of 
Eqs.(\ref{eq4B.30}-\ref{eq4B.39}) as said are given by the prescription
of Eq.(\ref{eq4B.24}).
This finishes our double insertions. We have not been able to 
evaluate all the two dimensional integrals into simple functions as we could
in section 4A.  In fact, for y=1, some of these integrals produce the 
Catalan number, an indication that they are probably related
to hyper-geometric functions. Our end result here
agrees with that in ref.\cite{Hoang3}.

\section{Numerical Results}

In this section, we present some numerical results. The mass in 
Eq.(\ref{eq2.1}) is
the b quark pole mass $M_b^{pole}$ 
and the charm quark mass used throughout is $\overline{\rm MS}$ mass $m_c(\bar{m}_c)$.
For our numerical work, we take

\begin{equation}
\bar{m}_c = 1.25 ~GeV, ~~~~~~
M^{pole}_b = 5.0 ~GeV.
\end{equation}
The typical energy scale involved is taken as
$\mu = {4\over 3} \alpha_s M^{pole}_b =
2.0 \ GeV$, at which the running coupling constant is evaluated to be

\begin{equation}
\alpha_s(2.0) = 0.30.
\end{equation}
These give $y=0.80.$  The energy of 1S state of bottomonium due to the
non-zero charm quark mass is shifted  by 

\begin{equation}
\Delta E^{NLO}_{1S} = -18.9 ~MeV, ~~~~~~
\Delta E^{NNLO}_{1S} = -48 ~MeV,
\end{equation}
due to next leading order and next to next leading order corrections, 
respectively.
Altogether, these amount to a shift of bottom quark 1S mass
\footnote{
There is a strong dependence on the scale $\mu$.  Our choice is 
$\mu =2.0 \ GeV.$  As a contrast, for 
$\bar m_c=1.5 \ GeV, \ M_b^{pole}= 5 \ GeV,$ and $\alpha_s=0.216$ 
at $\mu =4.7 \ GeV$, one 
would get $\Delta M_b^{1S}=-16MeV,$ close to what was obtained in
\cite{Hoang3}.  
Roughly speaking, our lower choice of $\mu$ enhances NLO by $9/4$ and 
NNLO by $27/8$ due to $\alpha_s.$
The dependence on $\mu$ for $\overline {MS}$ b mass is much weaker.
See \cite{Brambilla} for a discussion.
}.
\begin{equation}
\Delta M^{1S}_b = -33 \  MeV.
\end{equation}

With our explicit expressions for energy shift, we can calculate it for all
values of $y$, including  $y > 1$ .  We display in Fig. 3 
\begin{figure}
\center
\epsfig{file=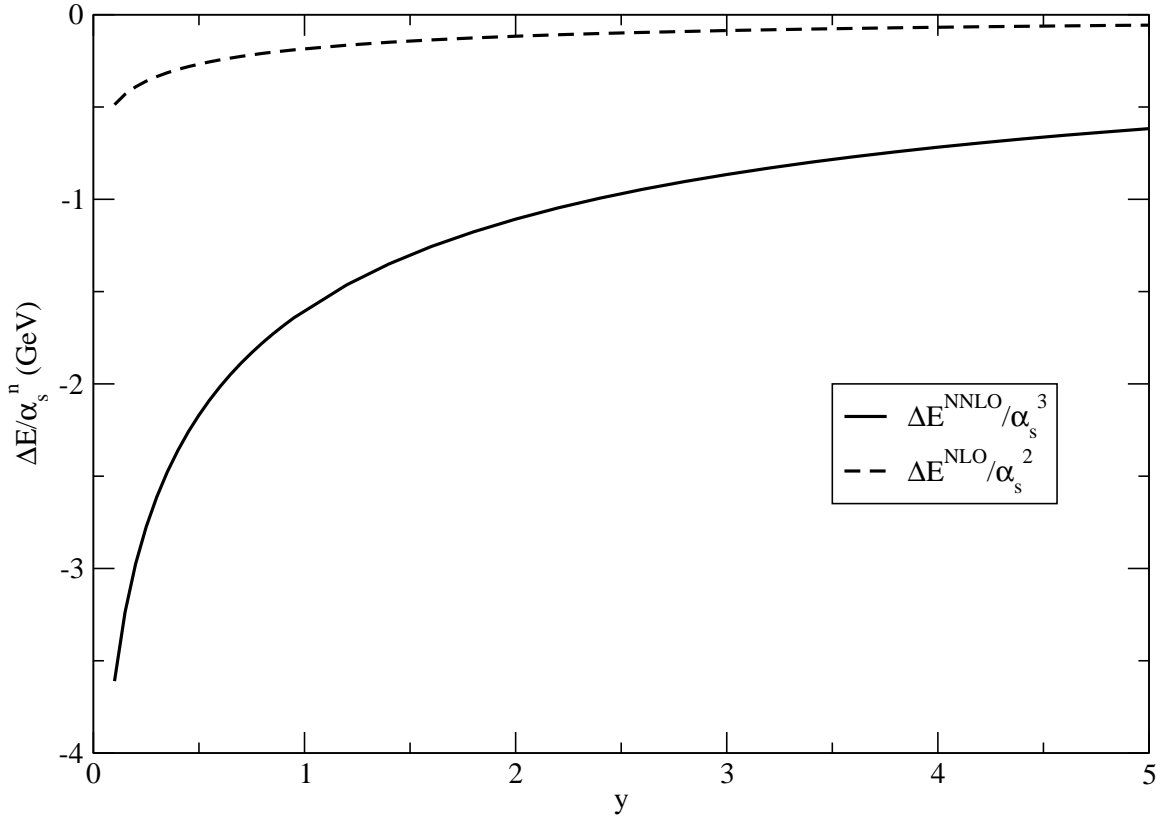,angle=-90,scale=0.65}
\caption{This figure compares the behaviors of $\Delta E^{NNLO}/\alpha_s^3$ 
against
$\Delta E^{NLO}/\alpha_s^2$ as functions of $y = \kappa_1/m_c$.
The solid line represents $\Delta E^{NNLO}/\alpha_s^3$, while the dashed
line represents $\Delta E^{NLO}/\alpha_s^2$.}
\end{figure}
the behavior of 
$\Delta E^{NNLO}/ \alpha_s^3$ and $\Delta E^{NLO}/ \alpha_s^2$ as functions 
of y to succinctly summarize the intricate interplay of masses, by 
eliminating their dependence on the running coupling
constant. In Fig. 4, 
\begin{figure}
\center
\epsfig{file=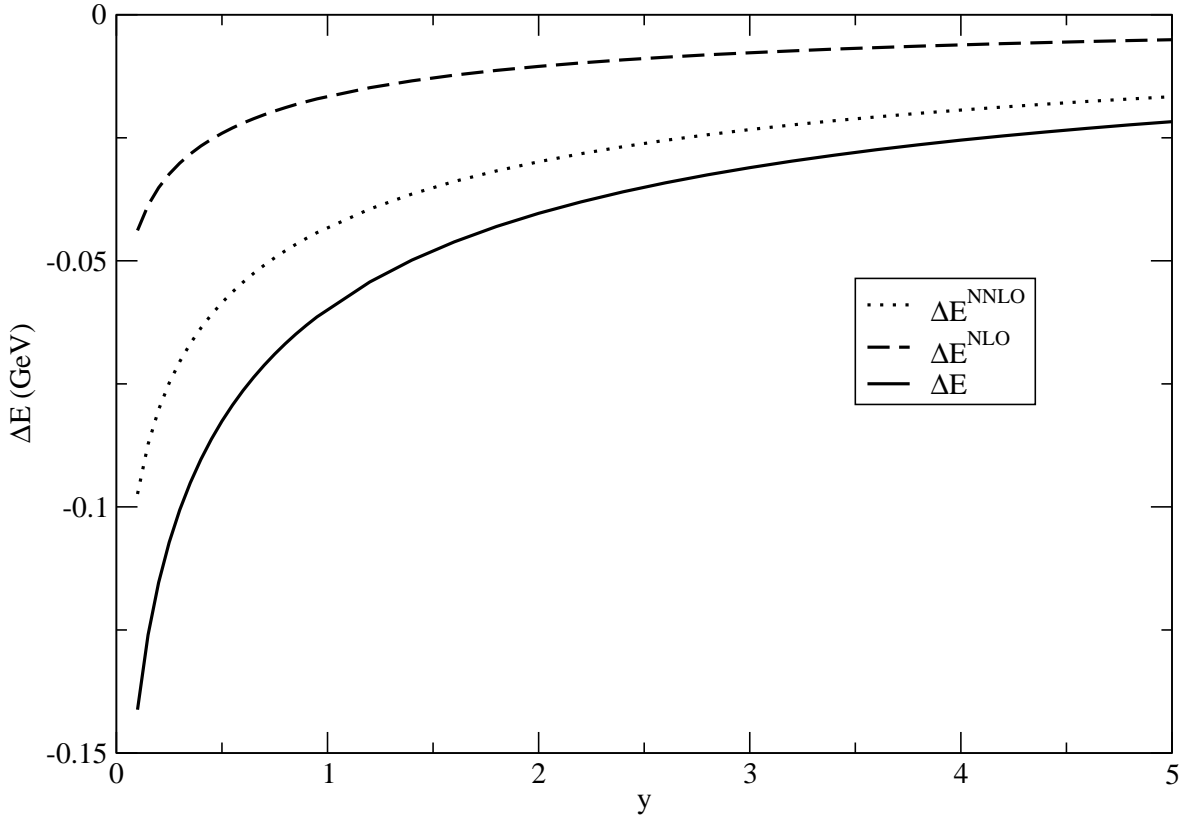,angle=-90,scale=0.65}
\caption{The 1S level shifts due to the charm mass effects are shown as 
functions of $y$ at NLO (dashed line) and NNLO (dotted line). 
The total contribution of NLO and NNLO
is represented by the solid line.
} 
\end{figure}
we plot the NLO, NNLO and the total contribution to 1S level energy shift as
functions of $y$.

Note that when $m_c \to 0$, which corresponds to $y\to \infty$, the
energy shift due
to double insertions tend to zero.  This point was emphasized by Hoang
\cite{Hoang3}, because the
leading order effects of the charm potential (Eq.(\ref{eq4A.12})) has no 
spatial dependence and
the orthogonality of the wave functions which make up the propagator
dictates this behavior.

The $t\bar{t}$ ``energy level shift'' due to
non-vanishing bottom quark mass can also be easily estimated.
We use
\begin{equation}
M_t =  174 \ GeV,  \ \ m_b =  4.2  \ GeV,  \ \
\alpha_s(M_t) = 0.108,
\end{equation}
which yield $y=2.98$.  Up to NNLO, the ``1S energy level shift'' due 
to the bottom quark mass effect reads
\begin{equation}
\Delta E = -25 \ MeV,
\end{equation}
which is suppressed by the smallness of the coupling constant as compared with
bottomonium.

\section{Concluding Remarks}

One motivation for this work is to give a formulation to
treat bound state and threshold physics completely in momentum space. 
For performing detailed high order calculations, it seems that the
momentum space may have a natural setting, because Feynman rules are
normally derived and given there. 

\bigskip
We have accomplished the construction of Coulomb wave functions
for all quantum numbers.  Our propagators are quite compact, which
among other things replace sums over principal quantum numbers by a
parametric integral.
The limit as the energy of the propagator(s) approaches any of the
bound state can be isolated easily.

\bigskip
We have applied this formalism to investigate the charm effects
on the 1S bottomonium level shifts to the next-next leading order.
Here, we have expressed all the single insertion results analytically
in elementary functions. For double insertions, where the ground state
is to be omitted as an intermediate state, it corresponds to a pole
subtraction in $1\over \epsilon$.

\bigskip
Clearly, our formalism is useful for other problems.  We have, for
example, cursorily looked into the shifts due to charm effects in
the next leading order on arbitrary bottomonium energy levels.
We find the needed technique to be a slight extension of
what we shall present in the appendix. Since the results for
the shifts are known \cite{Eiras1}, we shall not pursue it further.
The S-states due to a linear potential can also be easily
treated in this formulation.

\vglue0.5in
\noindent
{\bf \Large Acknowledgments }

\bigskip
The work by YPY is supported in part by the U. S. Department of Energy. The
work by HW is supported by the U.S. Department of Energy under
grant DE-FG02-91ER40681A29 (Task B).

\appendix
\section{Matrix Elements in Quantum Mechanics}
\bigskip
We want to show how some matrix elements in quantum mechanics can be 
reproduced in our approach.  Let us look at an integral for a fixed n
\begin{eqnarray}
\int d\xi \ {\tilde \gamma_l(\xi)\over \xi^s}{\tilde \gamma_l(\xi)\over \xi^t}
&=&
\int_{-\infty}^\infty dp \gamma_l^s(-p)\gamma^t_l(p)
\nonumber \\ &=& {A_l^2\over (i )^{s+t}}(-1)^{s+t}{1\over (s-1)!}{1\over (t-1)!} 
\int_{-\infty }^0 dp_1 \ p_1^{t-1}\int _{-\infty }^0dp_2 \ p_2^{s-1}
 \nonumber \\ & & \times\int _{-\infty}^\infty dp \ {(p_1+p+p_0)^{n-l-1}(p_2-p+p_0)^{n-l-1}
\over (p_1+p-p_0)^{n+l+1}(p_2-p-p_0)^{n+l+1}} .   
\end{eqnarray}
Using the residue theorem, we have

\begin{eqnarray}
\int d\xi \ {\tilde \gamma_l(\xi)\over \xi^s}{\tilde \gamma_l(\xi)\over \xi^t}
&=& {A_l^2\over (i )^{s+t}}(-1)^{s+t}{1\over (s-1)!}{1\over (t-1)!}
\int _{-\infty }^0 dp_1 \ p_1^{t-1}\int _{-\infty }^0dp_2 \ p_2^{s-1}\nonumber \\ & &
 \times {2\pi i\over (n+l)!}({d\over dx})^{n+l}{(2p_0+x)^{n-l-1}(p_1+p_2-x)
^{n-l-1}\over (p_1+p_2-2p_0-x)^{n+l+1}}|_{x=0} \ .
\end{eqnarray}

If $2(l+1)\ge s+t$, we can interchange the differentiation and the integrations 
over $p_{1,2}$.  Then, Eq.(\ref{eq2.34}) gives 
\begin{eqnarray}
\int d\xi \ {\tilde \gamma_l(\xi)\over \xi^s}{\tilde \gamma_l(\xi)\over \xi^t}
&=&{A_l^2\over (i )^{s+t}}{2\pi i\over (n+l)!}{1\over \Gamma(s+t)}({d\over dx})^{n+l}
b^{n-l-1}
\nonumber \\ & &\times\int_0^\infty dp'\ {p'^{s+t-1}(p'+x)^{n-l-1}\over (p'+b)^{n+l+1}}|_{x=0}
\nonumber \\ & =& (-1)^{s+t}{A_l^2\over (i )^{s+t}}{2\pi i\over (n+l)!}\sum_{i=0}^{n-l-1}
{(-2p_0)^{n-l-1-i}(n-l-1)!\over i!(n-l-1-i)!}\nonumber \\ &&
\times \sum_{j=0}^{s+t-1}{(-1)^{1+j}\over j!(s+t-1-j)!}{1\over n+l-i-j}
({d\over dx})^{n+l}b^{s+t-2l+i-2}|{x=0}, \nonumber \\
\label{eqA.3}
\end{eqnarray}
where $b=2p_0+x.$  Note that in most cases, only a small number of terms will 
survive the differentiation.  For example, we can check the normalization
$A_l$ by setting $s=t=l$, which leads to

\begin{equation}
\sum_{j=0}^{2l-1}{(-1)^{1+j}\over j!(2l-1-j)!}{1\over n+l-i-j}=
{1\over (n-i+l)(n-i+l-1)\dots(n-i-l+1)},
\end{equation}
and then 
\begin{eqnarray} 
\int _0^\infty d \xi \ u(\xi)^2 \ &=& \int_0^\infty d\xi \ 
{\tilde \gamma _l(\xi)\over \xi^l} \ {\tilde \gamma _l(\xi)\over \xi^l}
=\int_{-\infty}^\infty dp \ \gamma_l^l(p)\gamma_l^l(-p)
\nonumber \\ & =&{A_l^2\over (i)^{2l}} {2\pi i\over (n+l)!}
\sum_{i=0}^{n-l-1}(-2p_0)^{n-l-1-i}{(n-l-1)!(n-i-l)\over (n-i+l)!i!}
\nonumber \\ & &\ \ \  \times({d\over dx})^{n+l}b^{i-2}|_{x=0}.
\end{eqnarray}
Clearly only $i=0$ and $i=1$ contribute to the sum because of the d/dx 
operation.  We have after some simple algebra
 
\begin{eqnarray}
\int _0^\infty d \xi \ u(\xi)^2 \ &=& \int_0^\infty d\xi \ 
{\tilde \gamma _l(\xi)\over \xi^l} \ {\tilde \gamma _l(\xi)\over \xi^l}
\nonumber \\ & =&{A_l^2\over (i)^{2l}} ({1\over 2\kappa})^{2l+3}{2\pi\over (n+l)!}
(n-l-1)!(2n)\nonumber \\ &=&1,
\end{eqnarray}
as it must.

\bigskip
Next we consider $s+t=2l+1$, which will give us the expectation value of $1/\xi$.
This time we use in Eq.(\ref{eqA.3})

\begin{equation}
\sum_{j=0}^{2l}{(-1)^{1+j}\over j!(2l-j)!}{1\over n+l-i-j}=
{-1\over (n-i+l)(n-i+l-1)\dots(n-i-l)},
\end{equation}
which gives
\begin{eqnarray}
\int _0^\infty d \xi \ u(\xi){1\over \xi}u(\xi)  &=& \int_0^\infty d\xi \ 
{\tilde \gamma _l(\xi)\over \xi^l} {1\over \xi}{\tilde \gamma _l(\xi)\over \xi^l}
=\int_{-\infty}^\infty dp \ \gamma_l^{l+1}(p)\gamma_l^l(-p)
\nonumber \\ & =&{A_l^2\over (i)^{2l+1}} {2\pi i\over (n+l)!}
\sum_{i=0}^{n-l-1}(-2p_0)^{n-l-1-i}{(n-l-1)!\over (n-i+l)!i!}
\nonumber \\ & & \ \ \  \times({d\over dx})^{n+l}b^{i-1}|_{x=0}.
\end{eqnarray}
There is only one term i=0 in the sum which is not annihilated by d/dx and 
we come up with

\begin{eqnarray}
\int _0^\infty d \xi \ u(\xi){1\over \xi}u(\xi)  &=&
A_l^22\pi {(n-l-1)!\over (n+l)!}({1\over 2\kappa})^{2l+2}
\nonumber \\ &=&{1\over n^2 a},
\end{eqnarray}
which agrees with a well-known result.

\bigskip
As we stated earlier, to interchange the integration and the differentiation
operations in Eq.(\ref{eqA.3}), it is necessary that the condition $2(l+1)\ge s+t $ should 
be satisfied.  For the case $s+t=2l+2$, which is relevant for the expectation
value of $1/\xi^2$, we must treat it differently.  Thus, we write

\begin{equation}
\int_0^\infty d\xi \ {\tilde \gamma_l(\xi)\over \xi^{l+1}}
{\tilde \gamma_l(\xi)\over \xi^{l+1}}={A_l^2\over (i)^{2l+2}}
{2\pi i\over (n+l)!}{1\over (2l+1)!}B,
\end{equation}
where 
\begin{equation}
B=\int _0^\infty dp'  \ p'^{2l+1}({d\over dx})^{n+l}b^{n-l-1}
{(p'+x)^{n-l-1}\over (p'+b)^{n+l+1}}|_{x=0}.
\end{equation}
We rewrite
\begin{equation}
p'+x=p'+b-2p_0, \nonumber
\end{equation}
and expand in binomial the numerator
\begin{eqnarray} 
{(p'+x)^{n-l-1}\over (p'+b)^{n+l+1}} &=&{1\over (p'+b)^{2l+2}}
+{(n-l-1)(-2p_0)\over (p'+b)^{2l+3}}\nonumber \\ &  &
+{(n-l-1)(n-l-2)(-2p_0)^2\over 2(p'+b)^{2l+4}}+\dots.
\end{eqnarray}
Only the first term will be non-vanishing after the d/dx differentiations,
since all the others terms are convergent enough that we can interchange
the integration and differentiation operations and are proportional to
$(d/dx)^{n+l}b^k, \ k<n+l$.  Then effectively

\begin{equation}
B=\int _0^\infty dp' \ p'^{2l+1}({d\over dx})^{n+l}{b^{n-l-1}\over (p'+b)^{2l+2}}.
\end{equation}
After examining a few values of n and l, one can easily convince oneself that

\begin{equation}
B=-(n-l-1)!(2l)!b^{-(2l+1)},
\end{equation}
and thus

\begin{equation}
\int_0^\infty d\xi \ {\tilde \gamma_l(\xi)\over \xi^{l+1}}
{\tilde \gamma_l(\xi)\over \xi^{l+1}}={1\over n^3a^2}{1\over l+1/2},
\end{equation}
which again is a well-known result.

\end{document}